# New theory of femtosecond induced changes and nanopore formation

John Canning*[a], Matthieu Lancry[b], Kevin Cook[a], and Bertrand Poumellec[b]

[a] Interdisciplinary Photonic Laboratories (iPL), School of Chemistry, 222 Madsen Building F09, The University of Sydney, NSW 2006, Australia;
[b] LPCES/ICMMO, UMR CNRS-UPS 8182, Université Paris Sud 11, Bâtiment 415, 91405 Orsay Cedex, France
*john.canning@sydney.edu.au

## ABSTRACT

Recent results confirm the presence of molecular oxygen proving that recombination of dissociated silica bonds does not occur. This combined with the observation of nanopores within the nanograting structure in silica, leads to a new interpretation of femtosecond processing based on the unusual characteristics of quenching of tetrahedral silica compared to other glasses. This new approach suggests very different directions and implications for devices, including sensors, based on femtosecond laser processing of glasses.

## 1. INTRODUCTION

The impact of femtosecond laser processing on device processing, where temporal exposures can be much shorter than the normal phonon excitation and diffusive timescales, has been nothing short of remarkable. As well as demonstrating novel routes to many conventional phenomena, including waveguides [1-3] and fibre gratings [4-6], multiphoton ionisation has allowed many phenomena demonstrated with single photon processes to be highly localised within transparent media, opening up for the first time a practice a route to true 3-D laser processing. This extraordinary localisation has resulted in other variations of surface phenomena: for example, surface ripple gratings down to nano scale dimensions based on the Birnbaum effect [7] where optical light interferes with the generated plasma using picosecond or shorter wavelengths, has led to novel control of extended nanogratings inside a silica volume at any arbitrary position in 3-D [8,9]. Figure 1 shows an example of this where the writing beam (from an Yb$^{3+}$ doped femtosecond fibre laser) is either aligned along or orthogonal to the direction the beam is scanned in the glass. By controlling the laser polarisation these gratings can be controlled in orientation, suggesting the possibility of a novel 3-D holographic recording means. Structural color can also be created and tuned for different angles as shown in figure 2. In addition to more complex versions of known phenomena, the constraint of the material processing into a 3-D encapsulated volume leads to other novel effects not reported on the surface. This includes nanopore formation [3,10]. Figure 3 shows these nanopores in detail – they appear more like the mesostructure associated with zeolite materials than as individual pores formed in isolation. One of the outstanding issues, creating large uncertainty into the reliability and future control of these processes for real applications such as sensors, remains a satisfactory explanation of these phenomena – the observed nanograting formation [8] has led to the idea that plasma generation invokes breakdown of the Si-O bonds to release atomic $O^+$ and an electron, $e^-$. The coupling between the optical wavelength and the ultra short wavelength of the free electron and cumulative plasma, leads to short localisation of waves within the plasma (effectively "plasmon" modes). These can be nanoscale in dimension and are assumed to give rise to complex charge separation and different recombination both through diffusive processes and through electromotive gradient thrusts between high and low intensities. In this paper, we point out that irrespective of the details and the many complex variants, these models are mostly independent of the glassy state – this means the reported observations should be readily seen in any material regardless of the composition and details of processing required. But this is not the case and variations have only been observed for silica and relatively low concentration doped silica glasses [11]. We address this with an alternative model.

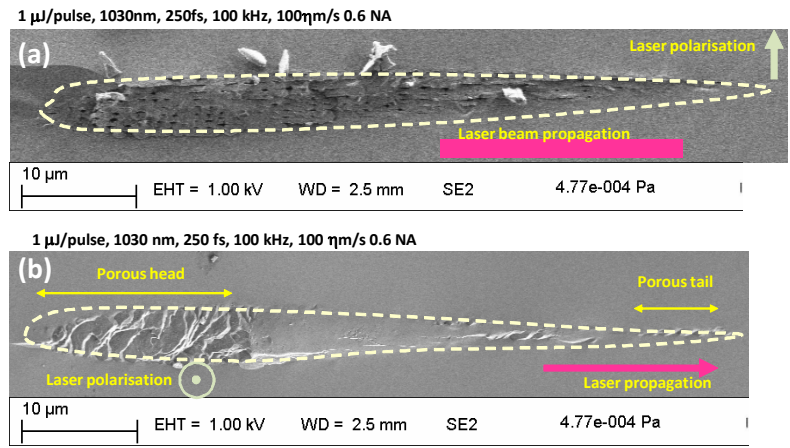

Figure 1. Induced damage trail across two waveguides: (a) laser polarisation is orthogonal to direction of travel of scanning laser beam (out of the pages) and (b) laser polarisation is aligned along the direction of scanning (out of the page).

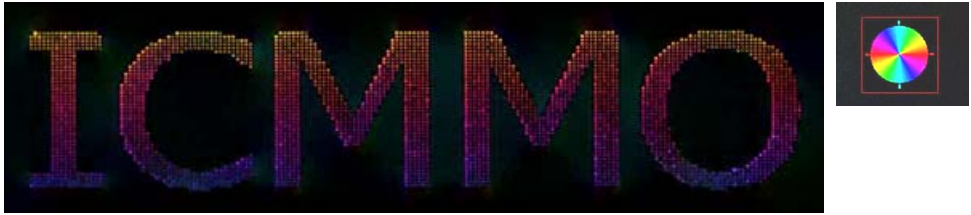

Figure 2. Structural colour by controlling nanograting orientation through control of the writing laser polarisation. Different tilt angles produce different colours at the one angle. The logo for the Institut de Chimie Moléculaire et des Matériaux d'Orsay is highlighted in varying colour with varying tilt angle (Fabricated by Lancry *et al*.).

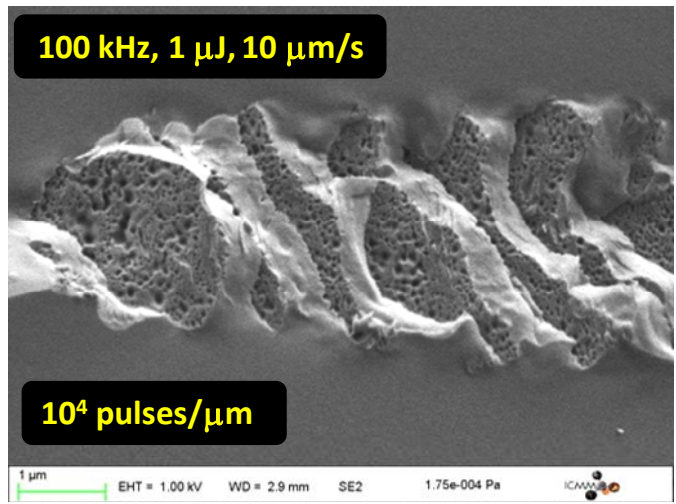

Figure 3. Close examination of sample nanopores found in the head of the nanograting structure of Figure 1 (b).

## 2. A SUMMARY OF THE FEMTOSECOND PROCESS

In figure 4 we summarise a route of formation which largely qualitatively describes much of what is accepted (at least in parts!). Assuming near infrared absorption (e.g. ~800 nm) into a glassy material such as silica where the band edge is below 180nm, the initial step obviously involves multiphoton ionisation (for 800nm ~ 6 photons is typically required for fused silica). This ionisation, when well above the ionisation threshold, leads to accelerated plasma formation and glass breakdown (or decomposition) as a result of highly energetic electrons themselves having sufficient energy to further ionise the glass. It also leads to a shockwave that can briefly expand the localised volume and densifies surrounding glass, enabling optical waveguides to support light below highly lossy damage regions of net reduced index [3]. What happens next during the cooling phase is not so clear given the different timescales involved with different processes. The generated electrons are sufficiently energetic to move away much more quickly (within 10 fs) than any induced atomic oxygen (within 100 fs to 1 ps). Since there is no significant phonon excitation or relaxation on these timescales (according to Eaton et al. [2,12] the local induced temperatures are < 3000K, restricting thermal diffusion times to > µs), normal thermal relaxation dynamics are unlikely to play a substantial role. An alternative model [10] suggests that given the high local field intensities associated with uneven plasma generation, especially at potential hotspots, electromotive forces are pushing charged species much more rapidly. Whilst these type of interactions may certainly be present (and in fact may very well play a role in the nanograting formation given the heterogeneity introduced by optical-plasma interference), they rely on an assumption of sufficient hot spot variation; yet the nanopores in Figure 3 show a fairly even distribution across most of the plates. Further, these fields would presumably drive recombination of Si-O, $O^+$ and $e^-$. The measured increase in glass $O_2$, both dissolved and in the nanopores [10], would indicate a significant amount of silica recombination is not occurring. One would have expected the same reasons why $O^+$ and $e^-$, despite differences in their migration times, would recombine will also mean full recombination to silica. But it appears not – on the other hand, thermal annealing at 1200 ºC does lead to full recovery of the glass network and the disappearance of these nanopores [10].

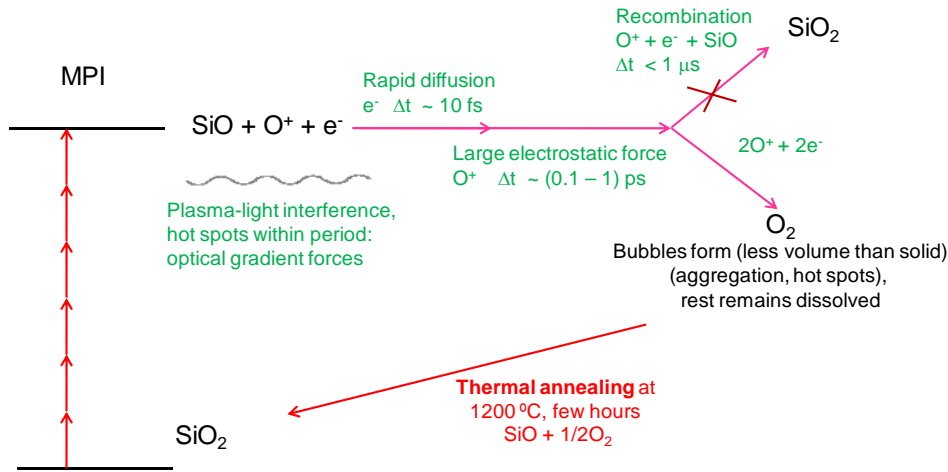

Figure 4. Summary of mechanisms involved with femtosecond processing of silica.

Therefore, there are two key observations of relevance:

(1) The induced glass changes are not inherently more stable than the initial starting phase. This contrasts, for example, with previous work on regenerated structures, where the induced changes by subsequent thermal processing of laser seeded structures alone can be made with a stability approaching that of the surrounding glass [13]. Other evidence of this is found in recent work reporting the successful thermal regeneration of femtosecond gratings to an even more stable temperature [14].

(2) Recombination back to silica is being prevented – the question is why?

It would seem there remain major challenges with existing interpretations in accounting for the observed changes and the presence of nanopores.

## 3. GLASS QUENCHING AND KINETICALLY DRIVEN RELAXATION

To explain the observations, we return to simple glass quenching of the amorphous state and suggest that the rapid deposition of heat through laser ionisation leads to an extreme in rapid cooling of a glass, where there is effectively (for the low repetition case) no overlap between excitation and quenching, an extremely unusual case that has no obvious analog in conventional glass processing by convective or radiative heat treatment alone. In the schematic of Figure 5 (a), an illustration of glass quenching of a typical glass former is depicted on a conventional *V-T* curve. As is expected from a simple quenching based on packing, the liquid state is usually less dense, and therefore lager volume, than the solid state, either amorphous or crystal, which has the optimal packing configuration. Representing the lowest free energy typically means a system, with sufficient time, drives towards the closest packing arrangement of its structure – for a random network this means the solid is denser than the liquid where local motions prevent densest packing, and that the crystal state is usually denser than the amorphous state.

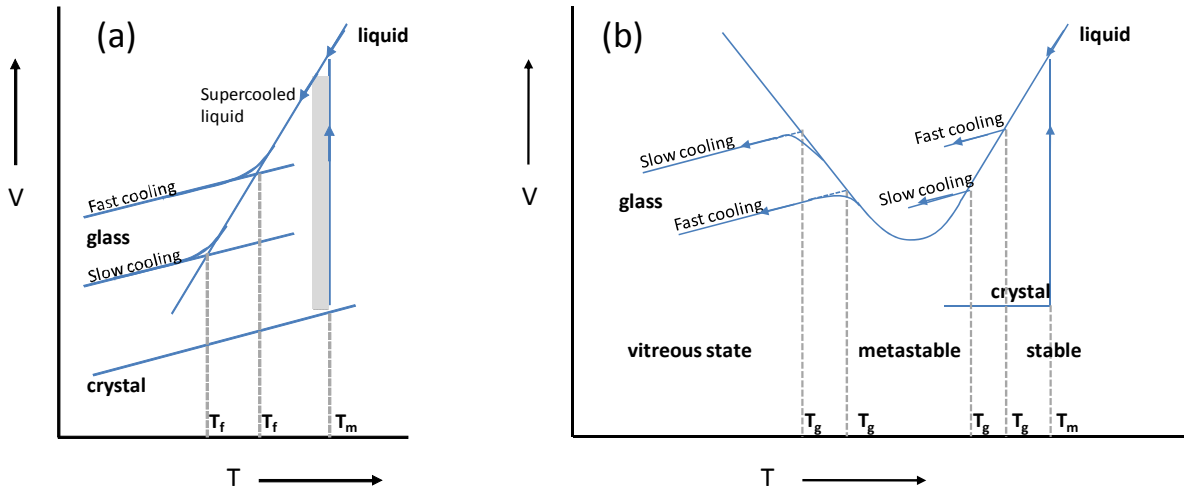

Figure 5. *V-T* curves for (a) typical glass forming liquids and for (b) low OH containing silica. Curves are derived from those found in [16].

However, silica is not a typical amorphous network – it is one of a handful of materials whose bond angles, and existing dipoles, present a serious theoretical challenge to a purely thermodynamic model of amorphous relaxation. In fact, the two most studied materials in science, and which still remain the most perplexing in many respects, have common tetrahedral structures and therefore related phenomena – silica and water. Both have the solid state volume larger than the liquid state – this can be explained as a topological restriction because the tetrahedral structure is short and compact and can only be packed in certain ways when bonds formed need to rigidly maintain the local angles. Therefore, the thermodynamic picture based on simple relaxation to the densest state is insufficient to explain these anomalies. A summary of the *V-T* diagram can be quite confusing to interpret – the version for silica is shown in Figure 5 (b) and is essentially that determined by Bruckner [15], and often used in text books [16]. If rapidly cooled from the molten state, crystallisation can be prevented and a metastable liquid/solid state (corresponding with the softened state of silica) is formed. This metastable state is associated with an unusually dense, and highly strained, liquid-like structure and further cooling leads to volume expansion rather than further contraction. In fact, in this cold regime we might expect kinetic relaxation to take over from thermodynamic relaxation. This leads to the unusual situation where rapid cooling leads to denser glass (freezing in of the liquid state before it has time to expand) than slower cooling.

Armed with the above review, we therefore propose the following: femtosecond laser processing is a relatively cold process where the starting phase is solid, and through ionisation, reaches almost instantly a quasi-metastable state that relaxes back to solid rapidly at relatively low temperatures. We may therefore (initially at least) consider the problem as being on the colder (left hand) side of the *V-T* diagram in Figure 5 (b) and can conclude that the final state of the glass is most likely to be denser and of lower molar volume than the starting phase, which was formed by conventional, slower thermal quenching from the metastable state. We can now begin to address the remainder of the problems raised above.

By having a smaller volume than the starting glass, a singular problem of silica is quickly realised: there is enormous constraint of the relaxation process by the surrounding glass. This constraint is equivalent to having enormous negative pressures, or tensile stresses, on the shrinking glass and is depicted schematically in Figure 6. It is intuitively obvious that there must be holes, cracks or nanopores formed for the quenching to occur. By deductive reasoning, and assuming a mainly homogenous environment overall, but noting that on a microscopic scale silica has great inhomogeneity, the lowest free energy of formation to facilitate quenching will be an even distribution of nanopores (or smallest features) – essentially, a mesostructured formation is expected. Any heterogeneity can complicate this and lead to tubes or nanocracks, (or possibly even periodic nanoplanes where there is a periodic heterogeneity) and other unusual clusters. [In other multicomponent glasses, heterogeneity is chemical in nature and rather than mesostructures forming, we would predict that strong phase separation occurs, even when the quenching is for typical glass behaviour since hydrostatic pressure is present as a result of a volume increase]. The question remains, however, is the difference in the final silica volume (density) such that the internal potential energy is sufficiently high that a porous gel-like (pure silica zeolite perhaps) network sufficient for this to occur in deference to the normal rigid tetrahedral structures? On average, when accommodating the volume of the nanopores, we estimate that the actual index is reduced substantially even if within the glass it has increased. This may also be aided by the expected volume expansion from the initial explosive plasma formation that led to densification in the surrounding regions.

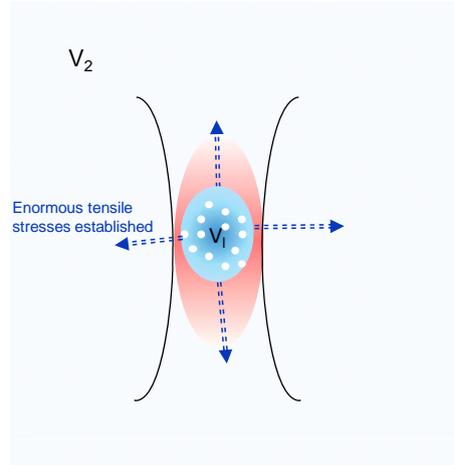

Figure 6. Schematic of the focal zone in a femtosecond laser processed glass. For typical glass-forming liquids, $V_1 > V_2$, since fast cooling leads to a less dense glass than the original slow cooled thermal quenched surrounding glass; but for silica $V_1 < V_2$, where fast cooling leads to a more dense glass than the surrounding slow cooled thermally quenched glass.

In fact, careful consideration reveals the actual situation with femtosecond processing is even more drastic than this: after photo-ionisation, the removal of oxygen immediately destroys the tetrahedral packing restraints and the glass relaxation in those regions where this has occurred is more akin to normal amorphous relaxation towards an *even denser state*. In fact, we postulate that this is the reason why recombination is no longer possible – the deviation from the tetrahedral coordination means there is a new, large potential energy barrier which is topological, or mechanical, in nature to recombination. This means neither $O^+$ nor $e^-$ can recombine with Si-O; the only solution is for them to eventually recombine with themselves to form $O_2$, as summarised in Figure 4. The denser relaxation of the network presumably means much of this $O_2$ is driven towards the nanopores; this in turn means $O_2$ formation is therefore unlikely to be the actual trigger for nanopores as has been suggested by other models, though it may play some role towards this. Thus each of the *V-T* diagrams alone is not addressing exactly this unusual situation; there may very well be a combination of processes occurring.

## 4. WHAT KIND OF SILICA?

We now have a new interpretation of femtosecond material processing, largely based on an intuitively more satisfying glass quenching model. Although we have started from typical glass quenching *V-T* diagrams, it's clear that the femtosecond process involves greater complexity since, in at least some of the glass, there may be a substantive transition occurring from a tetrahedral network to one which involves a structure closer to a non-even coordination such as $Si_5O_{10}$, for example. We may also go further – in [3] we pointed out that the large visible attenuation observed in the densified waveguiding region below the less dense mesostructured area appeared to be too significant to be explained by scattering alone; rather there are numerous defect sites (which is hardly surprising). We also noted that the attenuation has a characteristic profile that appears similar to the absorption profile of pure silicon – whilst presently speculative, further work is being undertaken to verify whether pure (most probably amorphous) Si can be generated by these mechanisms. Thus, the ionisation is not just of one Si-O silica bond on the Si but potentially more.

The expected coordination change, along with one outstanding observation – volume constraint - provides additional clues to its formation. As we noted above, volume constraint is another way of describing the presence of an enormous strain, or pressure, following plasma formation and relaxation. Thus we can summarise that the following general conditions are present: a relatively low temperature, fast quenching rates, and very high negative pressures (tensile stresses). Further, this occurs after the typical tetrahedral network dynamics may have been smashed – we therefore make a prediction based on these properties: the porous structure is a natural zeolite-like polyamorph of the glass responding to this pressure. In normal glasses the pressure is positive so it's unlikely to lead to tearing of the glass; rather the converse. [In fact we predict phase separation is possible in the case of multicomponent glasses where each component has different relaxation timescales – but that is another story]. The fabrication of mesostructured silica often involves self-assembly of structures from chemical precipitation or from a bottom-up nanoparticle approach. Laser induced mesostructures, with the incredible control associated using femtosecond lasers, is potentially a novel way to achieve such structures within volumes of solid materials.

Silica remains a key platform technology; thus the understanding of other processes induced by laser interactions in this particular material system remains extremely important. It is clear that much work remains to verify the model above; a starting point is to demonstrate the differences between those glasses that obey different quenching rules such as the two key ones in Figure 5. Nonetheless, the model already accounts for a number of observations without requiring any sophisticated pathways that are not well established. As result of this model, several predictions and directions can be made. A few examples include:

(1) It is less likely that any amorphous system that has an initial starting point where the glass is as dense, or denser, than the final state any of the phenomena thus far described can be observed. Therefore, the formation of nanopores should not be observable in traditional glassy materials.
(2) For many sensing and other applications, such as those relaying on waveguides and microfluidics using silica, low loss is extremely important. If the model is correct, we can predict that by adding dopants to the silica glass in sufficient quantity, we can avoid nanopores and other structural formations whilst still retaining fast turnaround in laser processing. This has immediate ramifications and directions for material research as well as alternative processing regimes to be explored.
(3) The thermal stabilisation of femtosecond processes could be very closely linked to the magnitude of this topologically driven relaxation – for higher temperature operation processing regimes and glass compositions that favor final relief of growing potential energies will lead to more stable changes.
(4) We propose laser processing, particularly with femtosecond lasers, as a new method for fabricating zeolite-like mesostructured materials to control the dielectric constant to an unprecedented level. This could have important implications for many sensor and device applications: for example, the next generation of computer chips where lower dielectric constant materials are increasingly difficult to achieve.

*Acknowledgements:* This work has been achieved in the frame of FLAG (Femtosecond Laser Application in Glasses) consortium project with the support of several organisations: the FP7 (IRSES-e-FLAG), the Agence Nationale pour la Recherche (ANR-09-BLAN-0172-01), the PRES UniverSud Paris (Pôle de Recherche et d'Enseignement Supérieur, 2008-39), the RTRA Triangle de la Physique (Réseau Thématique de Recherche Avancée, 2008-056T), and the Essonne administrative Department (ASTRE2007). Funding from Australian Research Council (ARC) DP 0879465 and 0770692 grants is acknowledged.